# EFFECT OF BLOCK MEDIUM PARAMETERS ON ENERGY DISSIPATION

K. X. Wang[a, b], N. I. Aleksandrova[c*], Y. S. Pan[a],

V. N. Oparin[c], L. M. Dou[b], and A. I. Chanyshev[c]

[a]Liaoning Technical University, Fuxin, 123000 China; kaixing_wang@163.com; panyish_cn@sina.com. [b]China University of Mining and Technology, Xuzhou, 221116 China; lmdou@126.com. [c]N. A. Chinakal Institute of Mining, Siberian Branch, Russian Academy of Sciences, Novosibirsk, 630091 Russia; *nialex@misd.ru; oparin@misd.ru; a.i.chanyshev@gmail.com. *Corresponding author.

**Abstract**: This paper describes energy distribution in a block medium simulated by a one-dimensional chain of masses joined by springs and dampers. Equations describing the motion of masses are solved by the methods of the theory of ordinary differential equations. The effect of the block medium parameters on energy dissipation is investigated. An approximate analytical solution is obtained that describes the total energy of a block medium at large values of time.

## INTRODUCTION

It is regarded in modern geomechanics and geophysics that rocks have a complex block hierarchical structure. According to this conception, a rock mass is a system of blocks embedded into each other and bound by interlayers of weakened or crack-like rocks [1]. Pendulum-type waves whose formation is induced by the geomechanical structure of a block medium was discovered in block medium dynamics [2, 3]. The propagation of pendulum waves in a block medium is widely studied both theoretically and experimentally [2-15]. Lately, the propagation of nonstationary waves in two- and three-dimensional block media has been investigated [12-15].

However, energy propagation in block media is understudied. It is noted in [4] that energy dissipation in block interlayers has a significant influence on wave propagation, so it is necessary to simulate this process theoretically. The interlayer viscosity is one of the reasons for energy dissipation. The viscous properties of the interlayers are accounted for in wave propagation models in a one-dimensional case for the medium that is a chain of blocks with pairs of elastic and damping elements, installed sequentially or parallel to each other [5]. The three-dimensional problem is described in [15].

This paper touches upon energy transfer and dissipation in a block medium on the basis of a one-dimensional model of a block medium with viscoelastic interlayers. Energy conversion and dissipation is comparatively analyzed in the case where the interlayer parameters and block mass vary in different regions of the block medium.

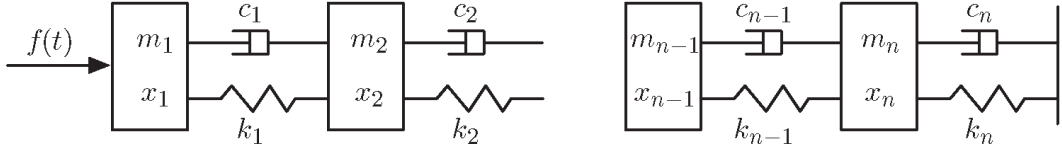

**Fig. 1.** Model of the block medium with viscoelastic interlayers.

## 1. FORMULATION OF THE PROBLEM AND SOLUTION

Figure 1 shows a block medium model as a chain of blocks. The interlayers between the blocks are simulated by elastic springs and viscous dampers. Note that wave propagation in block media is described rather accurately if the blocks are considered to be rigid undeformable solids.

An equation for the block medium dynamics is written in matrix form

$$M\ddot{x}(t) + C\dot{x}(t) + Kx(t) = F(t), \qquad (1)$$

where

$$M = diag[m_1, m_2, ..., m_n], \quad x(t) = [x_1(t), ..., x_n(t)]^T, \quad F(t) = [f(t), 0, ..., 0]^T,$$

$$C = \begin{bmatrix} c_1 & -c_1 & & & & & \\ -c_1 & (c_1+c_2) & -c_2 & & & & \\ & \ddots & \ddots & \ddots & & & \\ & & -c_{i-1} & (c_{i-1}+c_i) & -c_i & & \\ & & & \ddots & \ddots & \ddots & \\ & & & & & -c_{n-1} & (c_{n-1}+c_n) \end{bmatrix},$$

$$K = \begin{bmatrix} k_1 & -k_1 & & & & & \\ -k_1 & (k_1+k_2) & -k_2 & & & & \\ & \ddots & \ddots & \ddots & & & \\ & & -k_{i-1} & (k_{i-1}+k_i) & -k_i & & \\ & & & \ddots & \ddots & \ddots & \\ & & & & & -k_{n-1} & (k_{n-1}+k_n) \end{bmatrix},$$

$c_i$ and $k_i$ are the viscosity and rigidity of the $i$th interlayer, $m_i$ is the mass of the $i$th block, $x_i(t)$ is the displacement of the $i$th block; $i = 1, ..., n$; and $f(t)$ is the dynamic load.

Denotation $y(t) = [x(t), \dot{x}(t)]^T$ is introduced, and Eq. (1) is written as follows:

$$A\dot{y}(t) + By(t) = \tilde{f}(t). \qquad (2)$$

Here

$$A = \begin{bmatrix} C & M \\ M & 0 \end{bmatrix}, \quad B = \begin{bmatrix} K & 0 \\ 0 & -M \end{bmatrix}, \quad \tilde{f}(t) = \begin{bmatrix} F(t) \\ 0 \end{bmatrix}.$$

Shock load $f(t)$ is represented as the initial velocity $v$ of the first block, i.e., the initial conditions are set as:

$$x_i(0) = 0, (i = 1, ..., n), \quad \dot{x}_1(0) = v, \quad \dot{x}_i(0) = 0 \quad (i = 2, ..., n). \qquad (3)$$

The solution (2) with the initial conditions (3) has the form

$$y(t) = [x_1(t), ..., x_n(t), \dot{x}_1(t), ..., \dot{x}_n(t)]^T = \Phi d q_0, \qquad (4)$$

where $\Phi$ is the matrix comprised of eigenvectors $\varphi_i$ of the matrix $B^{-1}A$, i.e., $B^{-1}A\phi = \phi/\lambda$, $d = diag(e^{\lambda_1 t}, e^{\lambda_2 t}, ..., e^{\lambda_{2n} t})$, $\lambda_i$ denotes the characteristic values corresponding to $\varphi_i$, $q_0 = a^{-1}\Phi^T A y(0)$), $a = \Phi^T A \Phi = diag(a_1, a_2, ..., a_{2n})$ and $y(0)$ stands for the initial conditions.

Let the characteristic numbers of Eq. (2) be represented as $\lambda_r = -\beta_r + j\omega_r$, $\bar{\lambda}_r = -\beta_r - j\omega_r$ ($\beta_r > 0, \omega_r > 0$) and correspond to the eigenvectors $\varphi_r$ and $\bar{\varphi}_r$. Then, the displacement of the blocks is expressed via the vectors $\varphi_r$ and $\bar{\varphi}_r$ according to the equation

$$x(t) = \sum_{r=1}^{n} x_r(t), \tag{5}$$

where

$$x_r(t) = \phi_r e^{\lambda_r t} + \bar{\phi}_r e^{\bar{\lambda}_r t} = 2e^{-\beta_r t}\left[\operatorname{Re}(\phi_r)\cos\omega_r t - \operatorname{Im}(\bar{\phi}_r)\sin\omega_r t\right].$$

Expression (5) is used to determine the displacement and velocity of the $i$th block:

$$x_i(t) = \sum_{r=1}^{n} x_{ir} = \sum_{r=1}^{n}(\phi_{ir} e^{\lambda_r t} + \bar{\phi}_{ir} e^{\bar{\lambda}_r t}); \tag{6}$$

$$\dot{x}_i(t) = \sum_{r=1}^{n}(\lambda_r \phi_{ir} e^{\lambda_r t} + \bar{\lambda}_r \bar{\phi}_{ir} e^{\bar{\lambda}_r t}). \tag{7}$$

As the blocks are rigid, they have only kinetic energy in the dynamic perturbation propagation:

$$E_i = \frac{1}{2} m_i \dot{x}_i^2(t), \quad (i = 1, ..., n) \tag{8}$$

Using expressions (7) and (8), we write the equation for the kinetic energy of the block medium as

$$E_k = \sum_{i=1}^{n} E_i = \frac{1}{2}\sum_{i=1}^{n} m_i [\sum_{r=1}^{n}(\lambda_r \phi_{ir} e^{\lambda_r t} + \bar{\lambda}_r \bar{\phi}_{ir} e^{\bar{\lambda}_r t})]^2. \tag{9}$$

The potential energy of the block medium is determined by the stress state and interlayer deformation. Thus, the potential energy of the $i$th interlayer and block medium as a whole is determined from the expressions

$$U_i(t) = \frac{1}{2} k_i \Delta x_i^2, \quad E_p = \sum_{i=1}^{n} U_i(t) = \frac{1}{2}\sum_{i=1}^{n} k_i \left[\sum_{r=1}^{n}(\phi_{ir} e^{\lambda_r t} + \bar{\phi}_{ir} e^{\bar{\lambda}_r t} - \phi_{i+1,r} e^{\lambda_r t} - \bar{\phi}_{i+1,r} e^{\bar{\lambda}_r t})\right]^2, \tag{10}$$

where $\Delta x_i = x_{i+1} - x_i$ is the deformation of the $i$th interlayer between the blocks $x_i$ and $x_{i+1}$.

The total energy of the block medium is the sum of kinetic and potential energies

$$E = E_k + E_p \tag{11}$$

The Laplace time transform is applied to Eq. (1) in a partial case where $m_i = m$, $k_i = k$ and $c_i = c$ $(i = 1, ..., n)$. It is assumed that $p \to 0$ ($p$ is the parameter of the Laplace transform), which corresponds to $t \to \infty$. Thus, the following approximate solution describing the total energy of the block system for large time values is obtained:

$$E = \frac{mv^2}{2}\sqrt{\frac{m}{\pi c t}} = E_0 \sqrt{\frac{m}{\pi c t}}. \tag{12}$$

It follows from Eq. (12) that the total energy dissipation is directly proportional to the initial energy $E_0$ and the root of the block mass, and it is indirectly proportional to the root of viscosity of the interlayers and time.

## 2. NUMERICAL ANALYSIS OF ENERGY DISSIPATION IN A BLOCK MEDIUM

Figure 2 shows the conversion of the kinetic energy into potential energy and vice versa during the perturbation propagation. The kinetic energy $E_k$ and potential energy $E_p$ are calculated using Eqs. (4)-(10) with the following parameter values: $m_i = m = 10$ kg, $k_i = k = 6 \cdot 10^5$ kg/s$^2$, and $c_i = c = 35$ kg/s ($i = 1,...,20$). Next, these parameters are used as base parameters for the block medium. The dynamic interaction energy is 500 J with an initial condition $\dot{x}_1(0) = v = 10$ m/s.

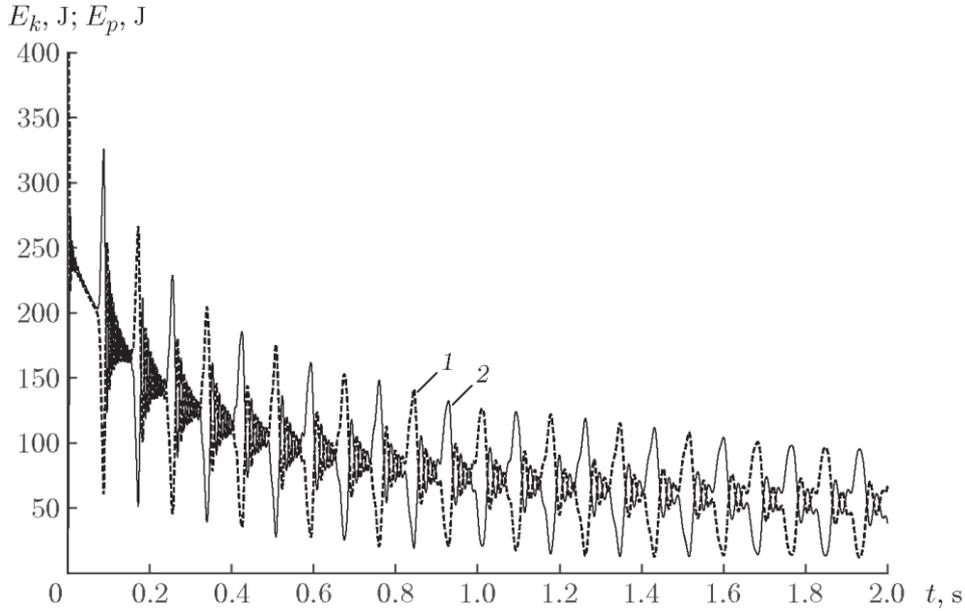

**Fig. 2.** Conversion of the kinetic energy (1) and potential energy (2) of the block medium.

Figure 2 shows both high- and low-frequency oscillations. The period of low-frequency oscillations induced by the reflection of waves from the boundaries $i = 1$ and $i = n$ ($n = 20$) can be calculated using $T_0 = n\sqrt{m/k}$ ($T_0 = 0,081$ s) [4], and the period of high-frequency oscillations is determined with the help of $T_1 = \pi\sqrt{m/k}$ ($T_1 = 0,012$ s). Figure 2 confirms the correctness of these analytical estimates.

Figure 3 illustrates the total energy dissipation in the block medium for different parameter values. The initial velocity $v$ is increased by a factor of $\sqrt{2}$ in order to keep the energy fed to the system at a value of 500 J in the case of a two-fold decrease in the mass. Figure 3 shows that the approximate solution (12) with a sufficient degree of accuracy describes the solution (4)-(11). According to the calculation results for the total energy of the system, its value is virtually independent of the interlayer rigidity in the case where the period of low-frequency oscillations of the system is significantly smaller than the time interval at which the process is studied, i.e., provided that $T_0 \ll T = 10$ s. The following estimate for the rigidity can be obtained from this inequality: $k \gg (n/T)^2 m$. As this inequality is fulfilled, the influence of the interlayer rigidity on the total energy of the system is low, so it can be calculated using the approximate solution (12).

Let the influence of the interlayer parameters and block mass in different regions of the block medium on the energy propagation and dissipation be analyzed. The block medium is split into three regions: region 1 is the first half of the block medium represented by blocks and interlayers with numbers $i = 1,...,10$, region 2 is the intermediate region comprised of blocks and interlayers with numbers $i = 6,...,15$, and region 3 is the second half of the block medium, namely blocks and interlayers with numbers $i = 11,...,20$.

Figure 4 shows the total energy dissipation in the block system for the case with a two-fold decrease in the interlayer viscosity in the regions 1-3 ($c_i = 17.5$ kg/s). Clearly, this slows down the total energy dissipation. The differences in the total energy in the cases with a two-fold viscosity in the regions 1-3 are significant only on the initial time interval $t < 4T_0$. With $t > 4T_0$, these differences can be ignored. With larger time values, the total energy can be sufficiently accurately described by the approximate solution (12) in which viscosity is equal to the mean value of the interlayer viscosity of the block medium: $c = c_*$. The calculations of the kinetic energy and potential energy in the block medium show that a two-fold decrease in the viscosity in the regions 1-3 increases the mean value equal to $E/2$, relative to which the kinetic energy and potential energy oscillate.

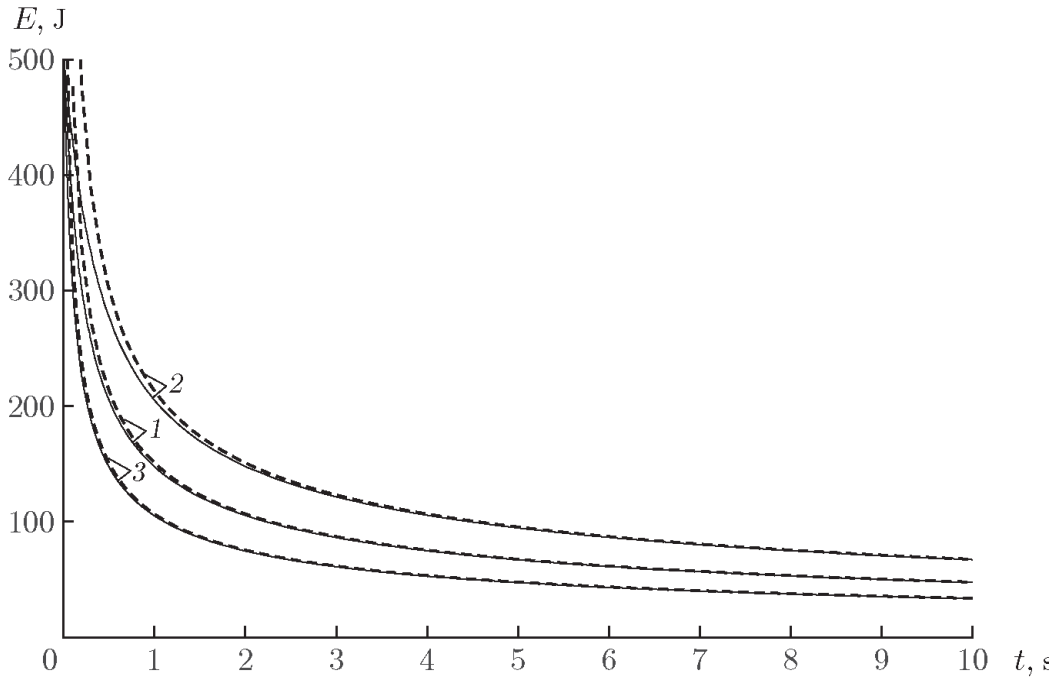

**Fig. 3.** Total energy dissipation of the block medium: the solid curves refer to the solution (4)-(11) and the dashed curves stand for the solution (12); (1) $m = 10$ kg, $k = 6 \cdot 10^5$ kg/s², and $c = 35$ kg/s; (2) $m = 10$ kg, $k = 6 \cdot 10^5$ kg/s², and $c = 17.5$ kg/s; (3) $m = 5$ kg, $k = 6 \cdot 10^5$ kg/s², and $c = 35$ kg/s.

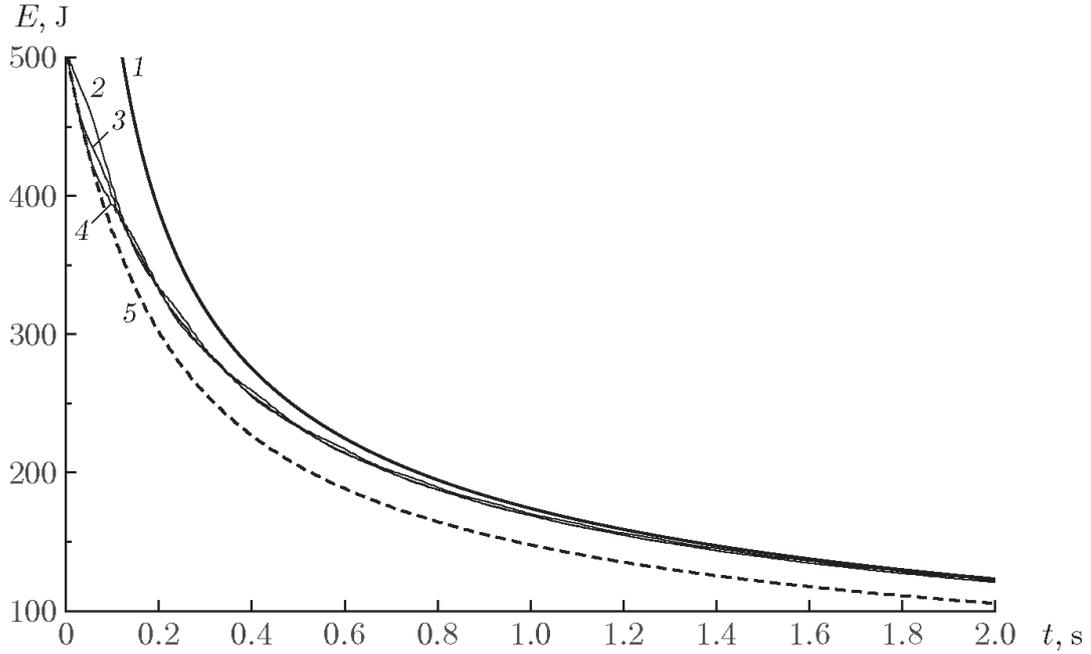

**Fig. 4.** Total energy dissipation with a two-fold decrease in the viscosity of the interlayers in different regions of the block medium: (1) solution (12) for $c_i = c_t = 26.25$ kg/s; (2) changes in the region 1; (3) changes in the region 2; (4) changes in the region 3; (5) energy calculated for the base parameters.

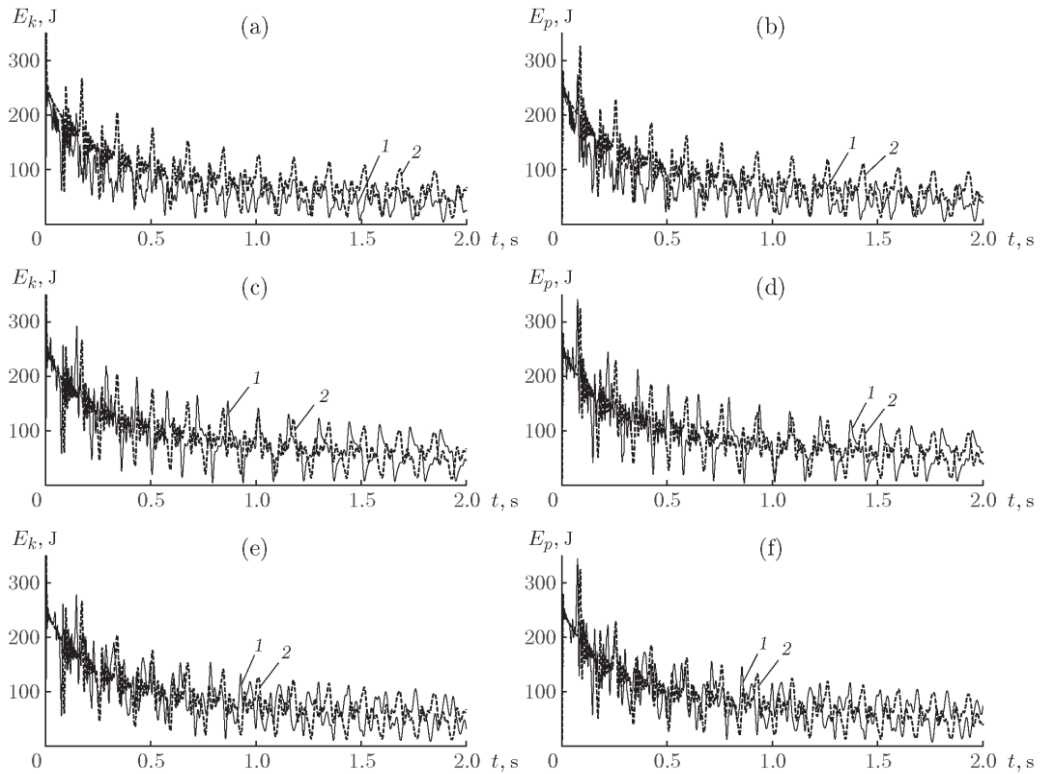

**Fig. 5.** Changes in the kinetic energy (a, c, and e) and potential energy (b, d, and f) of the block medium: (1) changes in the block mass in different regions; (2) original block medium; (a, b) changes in the block mass in the region 1; (c, d) changes in the block mass in the region 2; (e, f) changes in the block mass in the region 3.

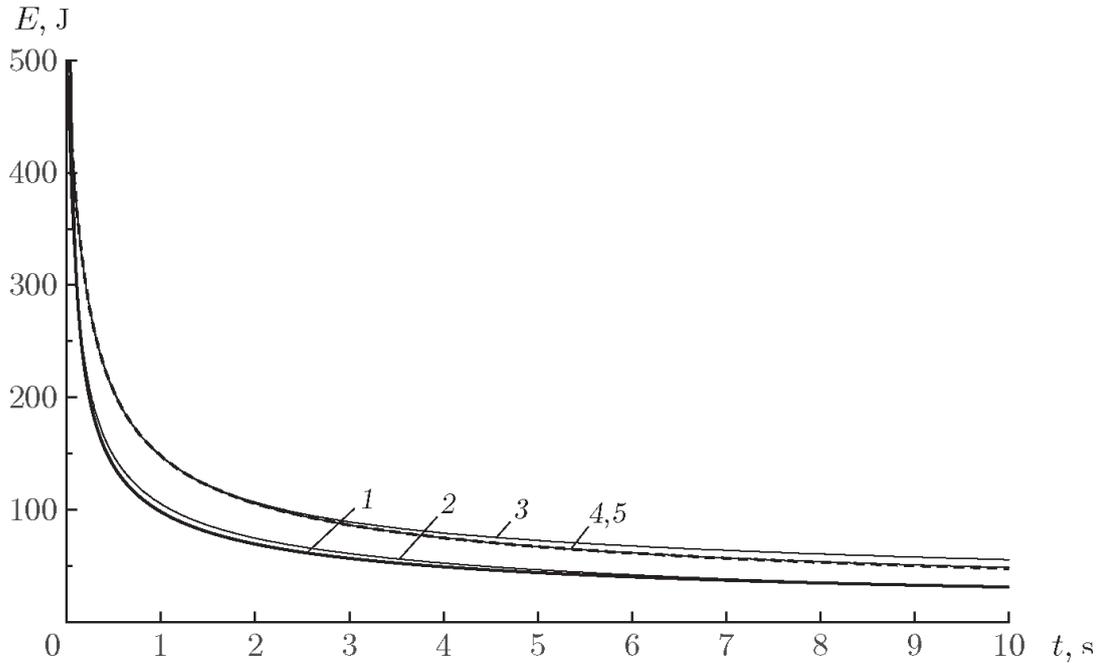

**Fig. 6.** Total energy dissipation with a two-fold decrease in the block mass in different regions of the block medium: (1) solution (12) obtained for $v = 10$ m/s, $m = m_t = 7.5$ kg, and the base values of the remaining parameters; (2) changes in the block mass in the region 1; (3) changes in the block mass in the region 2; (4) changes in the block mass in the region 3; (5) energy calculated for the base parameters.

Figure 5 shows changes in the kinetic energy and potential energy in the block medium with a two-fold decrease in the block mass m; in the regions 1-3 ($m_i = 5$ kg). The initial velocity $v$ is increased by a factor of $\sqrt{2}$ in order to keep the energy fed to the system at a value of 500 J in the case of a two-fold decrease in the block mass in the region 1. Figure 5 shows that, as the block mass drops in the region 1, the kinetic energy and the potential energy decrease along with the energy conversion period $T_0$. In the case of a decrease in the block mass in the regions 2 and 3, the energy conversion period $T_0$ becomes smaller, while the mean value relative to which the kinetic energy and potential energy oscillate, which is equal to half the total energy, remains virtually the same. In the case where the block mass in the regions 1 and 2 decreases, the oscillations of the kinetic energy and potential energy become more chaotic.

Figure 6 shows the total energy dissipation in the block system with a two-fold decrease in the block mass in different regions of the medium ($m_i = 5$ kg). Clearly, in the case of the reducing block mass in the region 1, the total energy dissipation becomes quicker, and the approximate solution (12) qualitatively and quantitatively describes the solution (4)-(11) quite accurately.

Figure 7 shows changes in the kinetic energy and potential energy with a two-fold decrease in the interlayer rigidity in the regions 1-3 ($k_i = 3 \cdot 10^5$ kg/s$^2$). As the interlayer rigidity decreases in the regions 1-3, the oscillations of the kinetic energy and potential energy become more chaotic. At the same time, the conversion period of the kinetic energy and potential energy $T_0$ becomes larger. Clearly, as the interlayer rigidity decreases in the regions 2 and 3, the oscillation amplitude of the kinetic energy and potential energy relative to the mean value significantly drops. However, it remains on the same level as in the original block medium in the case of a decrease in the interlayer rigidity in the region 1.

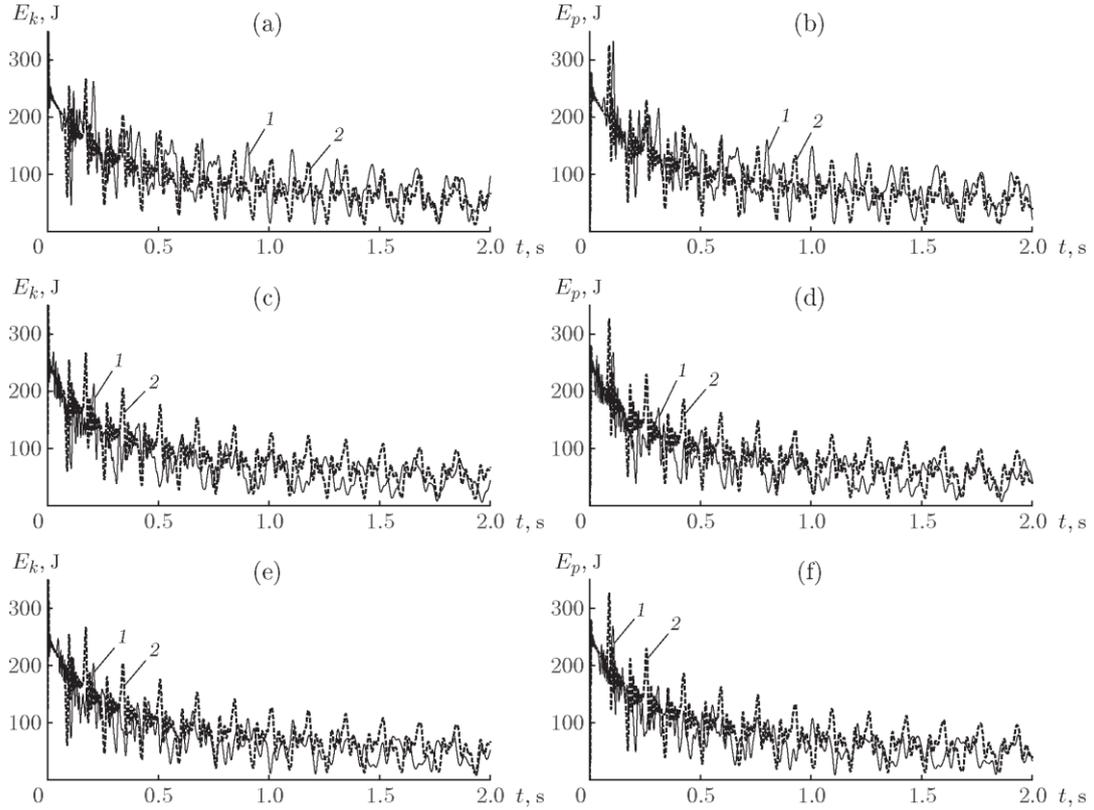

**Fig. 7.** Changes in the kinetic energy (a, c, and e) and potential energy (b, d, and f) of the block medium: (1) changes in the interlayer rigidity in different regions; (2) original block medium; (a, b) changes in the interlayer rigidity in the region 1; (c, d) changes in the interlayer rigidity in the region 2; (e, f) changes in the interlayer rigidity in the region 3.

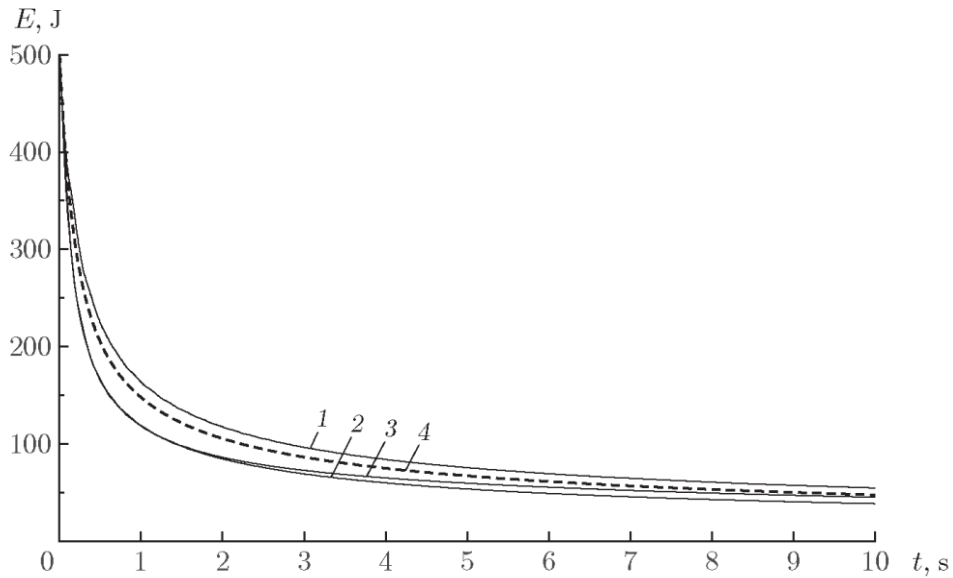

**Fig. 8.** Total energy dissipation with a two-fold decrease in the interlayer rigidity in different regions of the block medium: (1) changes in the interlayer rigidity in the region 1; (2) changes in the interlayer rigidity in the region 2; (3) changes in the interlayer rigidity in the region 3; (4) energy calculated for the base parameters.

Figure 8 shows the total energy dissipation with a two-fold decrease in the interlayer rigidity in the regions 1-3 ($k_i = 3 \cdot 10^5$ kg/s$^2$). Clearly, as the interlayer rigidity in the regions 1-3 decreases, the scattering of the total energy occurs more slowly in the original block medium. As the interlayer rigidity in the region 1 decreases, the total energy scatters more quickly.

**CONCLUSIONS**

The following conclusions can be made from this study:

As perturbations propagate in the block medium, the kinetic energy converts into potential energy and vice versa. The energy conversion period determined by the interlayer rigidity and block mass becomes larger with a decrease in the interlayer rigidity and smaller with a decrease in the block mass.

Energy dissipation in the block medium depends on the viscosity and rigidity of the interlayers and the block mass in different regions:

—the energy dissipation slows down as the interlayer rigidity decreases in the first half of the medium and becomes quicker as the interlayer rigidity in the middle section or second half of the medium decreases;

—as the block mass in the first half of the medium decreases, the energy dissipates more quickly, and a decrease in the block mass in the middle section or second half of the medium has no effect on the energy dissipation;

—as the interlayer viscosity decreases in different regions of the medium, the energy dissipation slows down in the block medium independently of the region where the viscosity varies.

**ACKNOWLEDGMENTS**

The work of Kaixing Wang, Yishan Pan, and Linming Dou is supported by the National Natural Science Foundation of China (Projects No. 51874163 and 51404129), China Postdoctor Foundation (Project No. 2017M611951), Open Projects of State Key Laboratory of Coal Resources and Safe Mining CUMT (Project No. SKLCRSM15KF05), and China Liaoning Province Natural Science Fund (Project No. 20170540429). N. I. Aleksandrova, V. N. Oparin, and A. I. Chanyshev's study was carried out within the framework of Fundamental Scientific Studies, Russia, no. gov. registration AAAA-A17-117122090002-5.